\documentclass[twocolumn,showpacs,preprintnumbers,amsmath,amssymb]{revtex4}

\usepackage{graphicx}
\usepackage{dcolumn}
\usepackage{bm}

\begin{document}

\preprint{APS}

\title{Big-Bang Nucleosynthesis verifies classical Maxwell-Boltzmann distribution}

\author{S.Q.~Hou$^{1,2}$}
\author{J.J.~He$^1$}
\email{jianjunhe@impcas.ac.cn}
\author{A.~Parikh$^{3,4}$}
\author{D.~Kahl$^5$}
\author{C.~Bertulani$^6$}
\affiliation{$^1$Key Laboratory of High Precision Nuclear Spectroscopy and Center for Nuclear Matter Science, Institute of Modern Physics, Chinese Academy of Sciences, Lanzhou 730000, China}%
\affiliation{$^2$University of Chinese Academy of Sciences, Beijing 100049, China}%
\affiliation{$^3$Departament de F\'{\i}sica i Enginyeria Nuclear, EUETIB, Universitat Polit\`{e}cnica de Catalunya, Barcelona E-08036, Spain}
\affiliation{$^4$Institut d'Estudis Espacials de Catalunya, Barcelona E-08034, Spain}
\affiliation{$^5$Center for Nuclear Study (CNS), the University of Tokyo, Wako Branch at RIKEN, 2-1 Hirosawa, Wako, Saitama 351-0198, Japan}
\affiliation{$^6$Texas A\&M University-Commerce, Commerce, TX 75429-3011, USA}%

\date{\today}

\begin{abstract}
We provide the most stringent constraint to date on possible deviations from the usually-assumed Maxwell-Boltzmann (MB) velocity distribution for nuclei in
the Big-Bang plasma. The impact of non-extensive Tsallis statistics on thermonuclear reaction rates involved in standard models of Big-Bang Nucleosynthesis (BBN)
has been investigated. We find that the non-extensive parameter $q$ may deviate by, at most, $|\delta q|$=6$\times$10$^{-4}$ from unity for BBN predictions
to be consistent with observed primordial abundances; $q$=1 represents the classical Boltzmann-Gibbs statistics. This constraint arises primarily from the 
{\em super}sensitivity of endothermic rates on the value of $q$, which is found for the first time. As such, the implications of non-extensive statistics in 
other astrophysical environments should be explored. This may offer new insight into the nucleosynthesis of heavy elements.
\end{abstract}



\maketitle
Big-Bang Nucleosynthesis (BBN) began when the universe was 3-minutes old and ended less than half an hour later when the nuclear reactions were quenched by the 
low temperature and density conditions in the expanding universe. Only the lightest nuclides (D, $^3$He, $^4$He, and $^7$Li) were synthesized in appreciable 
quantities through BBN, and these relics provide us a unique window on the early universe. Currently, standard BBN simulations give acceptable agreement between 
theoretical and observed abundances of D and $^4$He, but it is still difficult to reconcile the predicted $^7$Li abundance with the observation for the Galactic 
halo stars (GHS). The BBN model overestimates observations interpreted as primordial $^7$Li abundance by about a factor of three~\cite{Li2,Li}. This apparent 
discrepancy has promoted a wealth of experimental and theoretical inquiries. However, conventional nuclear physics seems unable to resolve the cosmological 
lithium problem (e.g., see Refs.~\cite{Cyburt08,boyd,kirsebom,wang,hamm,pizz}). Indeed the solution may lie in the refinement of observations given that the 
recently observed~\cite{SMC} $^7$Li abundance of the low-metallicity Small Magellanic Cloud (SMC) is consistent with the BBN predictions.

In the BBN model, the predominant nuclear-physics inputs are thermonuclear reaction rates (derived from cross sections). In the past decades, great efforts have
been undertaken to determine these data with high accuracy (e.g., see compilations~\cite{fcz67,Wagoner67,wagoner69,fcz75,cf88,Smiths93,Angulo99,Des04,Serpico04,yixu}). 
A key assumption in all thermonuclear rate determinations is that the velocities of ions may be described by the classical Maxwell-Boltzmann (MB) 
distribution~\cite{mandl08,Rolfs88}. It is well-known that the MB distribution was derived for describing the thermodynamic equilibrium properties of the ideal 
gas, where the particles move freely without interacting with one another, except for very brief collisions in which they exchange energy and momentum with each 
other or with their thermal environment. This classical distribution was ultimately verified by a high-resolution experiment~\cite{miller55} at temperatures 
around 900 K. However, in real gases, there are various effects (e.g., van der Waals interactions, relativistic speed limits, etc.) that make their speed 
distribution sometimes very different from the MB form. Moreover, stellar systems are generally subject to spatial long-range interactions, causing the 
thermodynamics of many-body self-gravitating systems to show some peculiar features differing drastically from typical ones~\cite{tar02}. Therefore it is worth 
asking: with what level of precision can the classical MB distribution be accurately applied to an extreme environment such as found in the Big-Bang plasma?

To address such issues, Tsallis proposed the concept of generalized non-extensive entropy and set up the non-extensive statistics or Tsallis 
statistics~\cite{Tsallis88,Gell04,Tsal09}. A parameter $q$ was introduced to describe the degree of non-extensivity of the system. $q$=1 represents the 
classical Boltzmann-Gibbs (BG) statistics; $q$$>$1 leads to an entropy decrease, providing a state of `higher order', whereas for $q$$<$1 the entropy increases 
as usual in a closed system, and the system can be considered to evolve towards `disorder'. The implications of deviations from classical statistics in stellar 
environments have been examined, in part, before. For example, Clayton et al.~\cite{Clyton75} explored using an ion distribution in the Sun with a depleted 
Maxwellian tail in an attempt to resolve the famous solar neutrino problem, although this was later understood to be a consequence of neutrino oscillation. 
Later, Degl'Innocenti et al.~\cite{heli98} derived strong constraints on such deviations by using the detailed helioseismic information of the solar structure, 
and found a small deviation $\delta q$ lying between (-1.0$\thicksim$0.4)\%. Recent work~\cite{Bertu13} has shown that only a small deviation $\delta q$ (lying 
between (-12$\thicksim$5)\% from the Maxwellian distribution is allowed for the BBN based on the Tsallis statistics. On the other hand, Torres 
et al.~\cite{torres} investigated the impact of non-extensive thermostatistics on the energy densities and weak interaction rates in the early Universe, and 
found $|\delta q|$$<$3.4$\times$10$^{-3}$.

In this Letter, we have numerically calculated the thermonuclear rates for relevant BBN reactions by using the non-extensive $q$-Gaussian distribution.
For the first time, the forward and reverse reaction rates have been obtained with such distribution coherently. With these non-extensive rates, the primordial
D, $^3$He and $^4$He and $^7$Li abundances are predicted by a BBN code with the up-to-date cosmological parameters and nuclear physics inputs. By comparing our 
predicted BBN abundances with the most recent astronomical observational data, we have tested the validity of describing the velocities of nuclei in a hot 
thermal plasma by the MB distribution and examined if the use of non-extensive statistics may help to resolve the cosmological lithium problem.

It is well-known that thermonuclear rate for a typical $1+2\rightarrow3+4$ reaction is usually calculated by folding the cross section $\sigma(E)_{12}$ with a 
MB distribution~\cite{Rolfs88}
\begin{equation}
\label{eq1}
\left\langle\sigma v\right\rangle_{12}=\sqrt{\frac{8}{\pi\mu_{12}(kT)^3}}\int_{0}^{\infty}\sigma(E)_{12}E\mathrm{exp}\left(-\frac{E}{kT}\right)\,dE,
\end{equation}
with $k$ the Boltzmann constant, $\mu_{12}$ the reduced mass of particles $1$ and $2$. In Tsallis statistics, the $q$-Gaussian velocity distribution can be 
expressed by~\cite{Silva2,Leubner04}
\begin{equation}
\label{eq2}
f_q(\mathbf{v})=B_q\left(\frac{m}{2\pi kT}\right)^{3/2}\left[1-(q-1)\frac{m\mathbf{v}^2}{2kT}\right]^{\frac{1}{q-1}},
\end{equation}
where $B_q$ denotes the $q$-dependent normalization constant. Thus, the non-extensive reaction rate becomes
\begin{widetext}
\begin{equation}
\label{eq3}
\left\langle\sigma v\right\rangle_{12}=B_q\sqrt{\frac{8}{\pi\mu_{12}}}\times\frac{1}{(kT)^{3/2}}\int_{0}^{E_\mathrm{max}}\sigma_{12}(E)E\left[1-(q-1)\frac{E}{kT}\right]^{\frac{1}{q-1}}\,dE,
\end{equation}
\end{widetext}
with $E_\mathrm{max}$=$\frac{kT}{q-1}$ for $q$$>$1, and +$\infty$ for 0$<$$q$$<$1. Here, the $q$$<$0 case is excluded according to the maximum-entropy 
principle~\cite{Tsallis88,Tsal09}. Usually, one defines the $1+2\rightarrow3+4$ reaction with positive $Q$ value as the forward reaction, the corresponding 
$3+4\rightarrow1+2$ with negative $Q$ value as the reverse one. Under the assumption of classical statistics, the ratio between reverse and forward rates can 
be expressed by~\cite{Rolfs88}
\begin{equation}
\label{eq4}
\frac{\left\langle\sigma v\right\rangle_{34}}{\left\langle\sigma v\right\rangle_{12}}=c\times\mathrm{exp}\left(-\frac{Q}{kT}\right),
\end{equation}
with a constant factor defined as $c=\frac{(2J_1+1)(2J_2+1)(1+\delta_{34})}{(2J_3+1)(2J_4+1)(1+\delta_{12})}\left(\frac{\mu_{12}}{\mu_{34}}\right)^{3/2}$.
With Tsallis statistics, however, the reverse rate is expressed as the following equation:
\begin{widetext}
\begin{equation}
\label{eq5}
\left\langle\sigma v\right\rangle_{34}=c\times B_q\sqrt{\frac{8}{\pi\mu_{12}}}\times\frac{1}{(kT)^{3/2}}\int_{0}^{E_\mathrm{max}-Q}\sigma_{12}(E)E\left[1-(q-1)\frac{E+Q}{kT}\right]^{\frac{1}{q-1}}\,dE.
\end{equation}
\end{widetext}
The previous work~\cite{Bertu13} determined forward rates using Eq.~\ref{eq3} but then simply determined reverse rates using Eq.~\ref{eq4}. In the 
present work we have used a coherent treatment and numerically calculated forward and reverse rates using Eqs.~\ref{eq3} and \ref{eq5}. In addition, the 
previous work~\cite{Bertu13} restricted the integral in Eq.~\ref{eq3} to a narrow energy range ($\pm$5$\Delta E_0$). This approximation is actually not 
sufficient for large values of $\delta q$~\cite{housq}. In the present work we have evaluated the integrals in Eqs.~\ref{eq3} and \ref{eq5} without such 
restrictions.

\begin{figure}[tbp]
\begin{center}
\includegraphics[width=8.5cm]{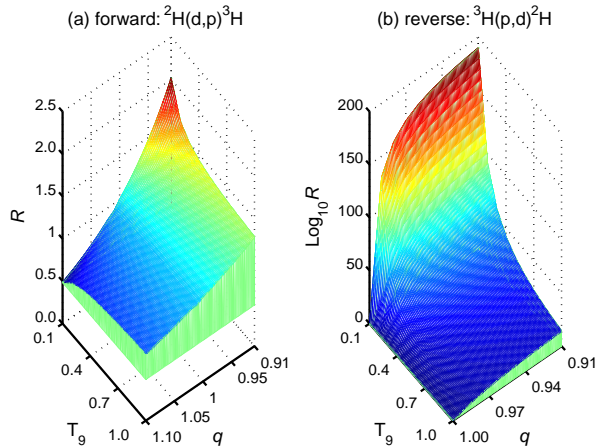}
\vspace{-3mm}
\caption{\label{fig1} (Color online) Ratio between rates calculated using Tsallis and MB distributions for the $^2$H(d,p)$^3$H reaction as functions of
temperature $T_9$ and $q$ values, (a) for forward reaction (in linear scale), and (b) for reverse reaction (in logarithmic scale).}
\end{center}
\end{figure}

\begin{figure}[tbp]
\begin{center}
\includegraphics[width=8.5cm]{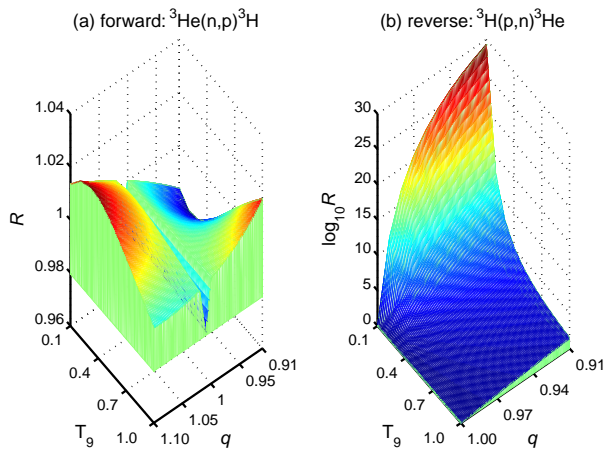}
\vspace{-3mm}
\caption{\label{fig2} (Color online) Results for the $^3$He(n,p)$^3$H reaction, see caption of Fig.~\ref{fig1}.}
\end{center}
\end{figure}

We show above the impact of $q$ values on the forward and reverse rates of two types of reactions. Here, $^{2}$H(d,p)$^{3}$H is taken as an example of the 
charged-particle-induced reaction, and $^3$He(n,p)$^3$H as that of the neutron-induced reaction. Both are among the most important reactions involved in the 
BBN. The ratios ($R$) between reaction rates determined with the Tsallis-distribution and MB-distribution are calculated for these two reactions. 
Figs.~\ref{fig1} and~\ref{fig2} show the results for forward and reverse rates as functions of temperature and $q$ value. Here, the cross section data for 
these two rates are taken from the compilations of Refs.~\cite{Angulo99,Des04}. In the region of 0.1$\leq$$T_9$$\leq$1.0 and 0.91$\leq$$q$$\leq$1.1, the 
forward rates calculated with the Tsallis-distribution deviate from the MB rates by relatively modest factors of, at most, 2 and 0.02 for the 
$^{2}$H(d,p)$^{3}$H and $^3$He(n,p)$^3$H reactions, respectively. However, the reverse rates for both types of reactions are {\em super}sensitive to deviations 
of $q$ from unity. For 0.91$\leq$$q$$\leq$1 (i.e., $q$$<$1), the corresponding Tsallis reverse rates deviate tremendously from the MB rates by about 200 and 
30 \textit{orders of magnitude} for $^{2}$H(d,p)$^{3}$H and $^3$He(n,p)$^3$H reactions, respectively. For instance, even with a very small deviation ($q$=0.999), 
the Tsallis reverse rate of $^{2}$H(d,p)$^{3}$H is about 10$^{10}$ times larger than the MB reverse rate at 0.2 GK. Here, the reverse rates with $q$$>$1 are 
not shown because they are negligible~\cite{housq} in comparison with the MB rates.

In order to explain qualitatively such {\em super}sensitivity, we define the factor [1-($q$-1)$\frac{E+Q}{kT}$]$^{\frac{1}{q-1}}$ in Eq.~\ref{eq5} as $P_q(E)$. 
If $|$1-$q$$|$$\ll$1, $P_q(E)$ can be expressed by the first-order approximation with~\cite{kan98}
\begin{equation}
\label{eq6}
P_q(E) \approx \mathrm{exp}\left[-\frac{E+Q}{kT}+\left(\frac{E+Q}{kT}\right)^2\times \frac{1-q}{2} \right].
\end{equation}
The ratio $R$ defined above for the reverse rates, can then be described approximately by $R>$ exp[($\frac{Q}{kT}$)$^2$$\times$$\frac{1-q}{2}$] for $q$$<$1, and
$R<$ exp[($\frac{Q}{kT}$)$^2$$\times$$\frac{1-q}{2}$] for $q$$>$1. It shows $R$ exponentially depends on the non-extensive parameter $q$, reaction $Q$ value
and temperature. In fact, the sensitivity of $P_q(E)$ (i.e., the tail of distribution) on $q$ results in the huge deviations for the reverse rates compared to 
those MB rates. The {\em super}sensitivity of reverse rate on the parameter $q$ has a very important impact: for $q$$<$1, the reverse rates are much larger than 
the forward rates, meaning that BBN is increasingly limited in its extent; on the other hand, for $q$$>$1, the reverse rates become negligible compared to the 
forward ones, an opposite effect to $q$$<$1.

We have investigated the impact of our new rates on BBN predicted abundances of D, $^3$He and $^4$He and $^7$Li using the code developed in Ref.~\cite{bbn}.
Recent values for cosmological parameters and nuclear physics quantities, such as the baryon-to-photon ratio $\eta$=(6.203$\pm$0.137)$\times$10$^{-10}$~\cite{WMAP9}, 
and the neutron lifetime $\tau_n=887.7$ s~\cite{nlife}, have been used in our model. The number of light neutrino families $N_{\nu}$=2.9840$\pm$0.0082 
determined by CERN LEP experiment~\cite{Nv06} supports the standard model prediction of $N_{\nu}$=3, which is adopted in the present calculation.
The reaction network involves nuclei with $A\leqslant9$ linked by the 34 reactions given in Table~\ref{tab1}. In total, 17 main reactions in the 
network~\cite{Smiths93} have been determined using non-extensive statistics, with 11 reactions~\cite{Smiths93} of primary importance and 6 of secondary
importance~\cite{Serpico04} in the primordial light-element nucleosynthesis. The standard MB rates~\cite{wagoner69,cf88,yixu,MFowler,ZHLi,Thomas93} have been 
adopted for the other reactions listed. Only those 11 primary important reactions were implemented with the non-extensive statistics in previous 
work~\cite{Bertu13}. Our predicted primordial BBN abundances with the usual MB distribution (i.e. $q$=1) are listed in Table~\ref{tab2}. The predictions by 
Bertulani et al.~\cite{Bertu13} (with MB distribution) and Coc. et al.~\cite{coc12}, as well as the up-to-date observed abundances are listed for comparison. 
Our results are quite consistent with those previous predictions~\cite{Bertu13,coc12,pizz}, and also agree well with the observations for D, $^3$He and $^4$He. 
In addition, all BBN predictions (with MB distribution) for the Li abundance are consistent with the value recently observed for the SMC~\cite{SMC}.

\begin{table}[t]
\centering
\setlength{\belowcaptionskip}{10pt}
\caption{\label{tab1} Nuclear reactions involved in the present BBN network. The non-extensive Tsallis distribution is implemented for 17 reactions shown in
bold face. The references for the nuclear physics data adopted for each case are also listed.}
\begin{tabular}{|ll|ll|}
\hline
Reaction & Ref. & Reaction & Ref. \\
\hline
(1) n $\to$ p     & \cite{nlife}                                       & (18) \textbf{$^2$H($\alpha,\gamma$)$^6$Li} & \cite{Angulo99,yixu}          \\
(2) $^3$H$\to^3$He & \cite{Tdecay}                                     & (19\footnotemark[1]) \textbf{$^3$H($\alpha,\gamma$)$^7$Li} & \cite{Des04}  \\
(3) $^8$Li$\to$2$^4$He & \cite{Tilley04}                               & (20\footnotemark[1]) \textbf{$^3$He($\alpha,\gamma$)$^7$Be} & \cite{Des04} \\
(4) $^6$He$\to^6$Li  & \cite{Tilley02}                                 & (21\footnotemark[1]) \textbf{$^2$H(d,n)$^3$He} & \cite{Des04}              \\
(5) $^6$Li(n,$\gamma$)$^7$Li & \cite{MFowler}                          & (22\footnotemark[1]) \textbf{$^2$H(d,p)$^3$H} & \cite{Des04}               \\
(6) $^2$H(n,$\gamma$)$^3$H   & \cite{wagoner69}                        & (23\footnotemark[1]) \textbf{$^3$H(d,n)$^4$He} & \cite{Des04}              \\
(7) $^6$Li(p,$\gamma$)$^7$Be & \cite{yixu}                             & (24\footnotemark[1]) \textbf{$^3$He(d,p)$^4$He} & \cite{Des04}             \\
(8) $^6$Li(n,$\alpha$)$^3$H  & \cite{cf88}                             & (25) \textbf{$^7$Be(d,p)2$^4$He} & \cite{Paker72,cf88}                     \\
(9) $^3$He(n,$\gamma$)$^4$He  & \cite{wagoner69}                       & (26) \textbf{$^7$Li(d,n)2$^4$He} & \cite{cf88}                             \\
(10\footnotemark[1]) \textbf{$^1$H(n,$\gamma$)$^2$H}  & \cite{npD}     & (27) $^3$He($^3$He,2p)$^4$He & \cite{cf88}                                 \\
(11\footnotemark[1]) \textbf{$^3$He(n,p)$^3$H}    & \cite{Des04}       & (28) $^7$Li(n,$\gamma$)$^8$Li & \cite{wagoner69}                           \\
(12\footnotemark[1]) \textbf{$^7$Be(n,p)$^7$Li}    & \cite{Des04}      & (29) $^9$Be(p,$\alpha$)$^6$Li & \cite{cf88}                                \\
(13\footnotemark[1]) \textbf{$^7$Li(p,$\alpha$)$^4$He} & \cite{Des04}  & (30)  2$^4$He(n,$\gamma$)$^9$Be & \cite{cf88}                              \\
(14\footnotemark[1]) \textbf{$^2$H(p,$\gamma$)$^3$He}  & \cite{Des04}  & (31) $^8$Li(p,n)2$^4$He & \cite{wagoner69}                                 \\
(15) \textbf{$^3$H(p,$\gamma$)$^4$He}  & \cite{PTHe4}                  & (32) $^9$Be(p,d)2$^4$He & \cite{cf88}                                      \\
(16) \textbf{$^6$Li(p,$\alpha$)$^3$He} & \cite{Angulo99,yixu}          & (33) $^8$Li(n,$\gamma$)$^9$Li & \cite{ZHLi}                                \\
(17) \textbf{$^7$Be(n,$\alpha$)$^4$He} & \cite{King77}                 & (34) $^9$Li(p,$\alpha$)$^6$He & \cite{Thomas93}                            \\
\hline
\end{tabular}
\footnotetext[1]{Of primary importance in the primordial BBN~\cite{Smiths93}.}
\end{table}

\begin{table}[tbp]
\centering
\setlength{\belowcaptionskip}{10pt}
\caption{\label{tab2} The predicted abundances for the BBN primordial light elements with the usual MB distribution (i.e. $q$=1). The observational data
are listed for comparison.}
\begin{tabular}{|c|c|c|c|c|}
\hline
Abundance & Present & Ref.~\cite{Bertu13} & Ref.~\cite{coc12} & Observation \\
\hline
$^4$He                       & 0.2485 & 0.249 & 0.2476 & 0.2465$\pm$0.0097~\cite{aver13} \\
D/H($\times$10$^{-5}$)       & 2.54   & 2.62  & 2.59   & 2.53$\pm$0.04~\cite{cooke14} \\
$^3$He/H($\times$10$^{-5}$)  & 1.01   & 0.98  & 1.04   & 1.1$\pm$0.2~\cite{bania02} \\
$^7$Li/H($\times$10$^{-10}$)\footnote{A value of $^7$Li/H=(4.8$\pm$1.8)$\times$10$^{-10}$ was recently observed for the SMC~\cite{SMC}.} & 5.34   & 4.39 & 5.24   & 1.58$\pm$0.31~\cite{Li2} \\
\hline
\end{tabular}
\end{table}

\begin{figure}[tbp]
\begin{center}
\includegraphics[width=7.0cm]{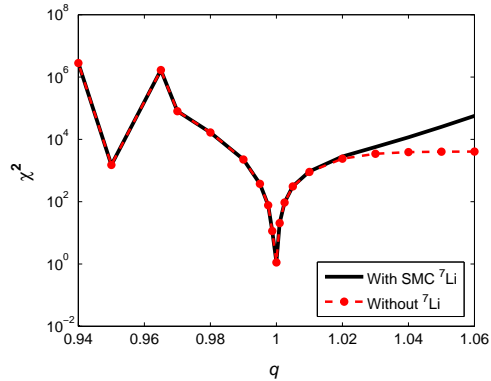}
\vspace{-3mm}
\caption{\label{fig3} (Color online) Calculated $\chi^2$ as a function of non-extensive parameter $q$ by using the observed primordial abundance data. The (red) 
dashed curve (points) represents the case using the observed primordial abundances of D, $^3$He and $^4$He to determine $\chi^2$, and the black solid curve 
(points) for the case with the additional constraint of the observed SMC $^7$Li abundance.}
\end{center}
\end{figure}

The least-squares fits have been performed to search for an appropriate $q$ value with which one can well reproduce the observed primordial abundances. 
The $\chi^2$ is defined by the minimization of
\begin{equation}
\label{eq:fit}
\chi^2=\sum_{i}\left[\frac{Y_i(q)-Y_i(obs)}{\sigma_i}\right]^2,
\end{equation}
where $Y_i$($q$) is the abundance (of nuclide $i$) predicted with a non-extensive parameter $q$, and $Y_i(obs)$ is the observed one with $\sigma_i$ the
observational error. We have examined two cases: the first calculates $\chi^2$ using only the primordial D, $^3$He and $^4$He abundances, and the second
uses the SMC $^7$Li abundance as well. $\chi^2$ is plotted in Fig.~\ref{fig3} as a function of parameter $q$ varying from 0.94 to 1.06 (i.e., deviation of
$\pm$6\%). The relatively narrower range of $q$ explored here, compared to Ref.~\cite{Bertu13}, was chosen owing to the {\em super}sensitivity of the reverse
rates on $q$ as discussed above. This {\em super}sensitivity has a very important consequence: the Tsallis distribution cannot be allowed to deviate very much
from the classical MB distribution for the Big-Bang plasma. Figure~\ref{fig3} shows gracefully that the $\chi^2$ function is minimized at unity (i.e., $q$=1). 
The predicted D/H, $^3$He/H and $^4$He abundances agree with observations at the 1$\sigma$ level for deviations $\delta q$=(-0.06$\thicksim$0.001)\%, 
(-0.8$\thicksim$0.5)\% and (-0.2$\thicksim$0.2)\%, respectively. If we adopt the strongest observational constraint, i.e., that of the deuteron, we conclude 
that the deviation from the MB distribution must be less than $|\delta q|$=6$\times$10$^{-4}$. Comparing this value to the previously estimated deviations of 
(-0.34$\thicksim$0.34)\%~\cite{torres} and (-12$\thicksim$5)\%~\cite{Bertu13}, our constraint is the most stringent one to date. For the fit with the SMC $^7$Li 
abundance (labeled ``with $^7$Li" in Fig.~\ref{fig3}), the predicted $^7$Li/H abundance agrees with observations at the 1$\sigma$ level for deviations 
$\delta q$=(-0.4$\thicksim$0.3)\%. 

In addition, we have searched for a value of $\delta q$ that would reconcile BBN predictions with the $^7$Li abundance observed in Galactic halo stars: we find 
$\delta q$$\approx$0.985\%. Unfortunately, even this small deviation of $q$ from unity would make the predicted abundances of D, $^3$He, and $^4$He significantly 
deviating from the observations by 43.5$\sigma$, 1.6$\sigma$ and 3.0$\sigma$, respectively. Therefore, the cosmological lithium problem cannot be solved by the 
application of non-extensive statistics discussed here.

We find that the comparison of predicted and observed primordial abundances verifies the applicability of the classical Maxwell-Boltzmann distribution for the 
velocities of nuclei during Big-Bang nucleosynthesis. Nonetheless, the present work reveals the striking impact that the use of non-extensive statistics may 
have on calculations of thermonuclear reaction rates. Indeed, nucleosynthesis in more extreme astrophysical sites such as supernova explosions may be profoundly 
affected by the {\em super}sensitivity of endothermic rates on the value of the non-extensive parameter $q$. We encourage extensions of the present study to 
further interrogate and test the usual assumptions of classical statistics in stellar environments.

\begin{acknowledgments}
This work was financially supported by the Major State Basic Research Development Program of China (2013CB834406) and the National Natural Science Foundation of
China (Nos. 11135005, 11321064). AP was supported by the Spanish MICINN (Nos. AYA2010-15685, EUI2009-04167), by the E.U. FEDER funds as well as by the ESF 
EUROCORES Program EuroGENESIS. CB acknowledges support under U.S. DOE Grant DDE- FG02- 08ER41533, U.S. NSF grant PHY-1415656, and the NSF CUSTIPEN grant
DE-FG02-13ER42025. JJ would like to express appreciation to Taka Kajino (NAOJ, Tokyo) and Diego F. Torres (IEEC-CSIC, Barcelona) who made helpful comments on 
the manuscript.
\end{acknowledgments}

\end{document}